\documentstyle[12pt]{article}
\newcommand{\doublespace}{\renewcommand{\baselinestretch}{1.75}
\Large\normalsize}
\begin{document}
\doublespace
\begin{titlepage}

\centerline{\bf GRAVITATIONAL ENERGY OF CONICAL DEFECTS}
\vskip 2.0cm
\centerline{Jos\'e W. Maluf$^*$ and Andreas Kneip}
\centerline{Universidade de Bras\'ilia}
\centerline{C.P. 04385}
\centerline{70.919-970 Bras\'ilia DF}
\centerline{Brazil}
\bigskip

\date{}

\begin{abstract}
The energy density $\varepsilon_g$ of asymptoticaly flat gravitational 
fields can be calculated 
from a simple expression involving the trace of the torsion tensor. 
The integral of this energy density over the whole space yields 
the ADM energy. Such energy expression 
can be justified within the framework of the teleparallel equivalent of
general relativity, which is an alternative geometrical formulation of
Einstein's general relativity. In this letter we apply $\varepsilon_g$
to the evaluation of the energy per unit length of a class of 
conical defects of topological nature, which
include disclinations and dislocations (in the terminology of
crystallography). Disclinations correspond to cosmic strings, and for a
spacetime endowed with only such a defect we obtain precisely
the well known expression of energy per unit length.
However for a pure spacetime dislocation the total gravitational
energy is zero.
\end{abstract}
\thispagestyle{empty}
\vfill
\noindent PACS numbers: 04.20.Cv, 04.20.Fy, 98.80.Cq\par
\noindent (*) e-mail: wadih@guarany.cpd.unb.br
\end{titlepage}
\newpage

\noindent {\bf I. Introduction}\par
\bigskip
\noindent It is believed that phase transitions in the early
universe can give rise to topological defects, which can lead to
very important cosmological consequences\cite{Vilenkin}. 
In order to understand the formation of galaxies
and cluster of galaxies it has been suggested that these structures
have evolved from the gravitational instability of small density
fluctuations. One of the major unresolved cosmological problems is
the origin of these initial density fluctuations. One possibility is
that the latter are due to cosmic strings\cite{Zeldovich,AVilenkin}.
For this reason cosmic strings have been widely studied in the
literature.

In the terminology of crystallography\cite{Nabarro,Rivier}
cosmic strings correspond  
to disclinations, which is one possible defect in a crystal. 
To our knowledge, other commom crystal defects like dislocations 
have not yet been considered in the various cosmological models. 
From a geometrical point of view, disclinations and dislocations
are conical singularities in a flat, four dimensional Lorentzian
spacetime, i.e., they can be described by a metric which is flat 
away from the $r=0$ axis, but with a coordinate singularity that
cannot be removed. Tod\cite{Tod} has recently considered these  
defects in a unified fashion. He generalized an argument 
due to Vickers\cite{Vickers}, according to which a conical
singularity like a cosmic string can be interpreted in terms of
a $\delta$-function of curvature supported in the origin of the
$(x,y)$ plane, say. Tod extended this idea to conical sigularities
of the dislocation type and argued likewise that instead of
interpreting the latter as defects in an otherwise 
flat Minkowski spacetime, one can alternatively consider a flat 
Minkowski spacetime endowed with a delta function of torsion 
as a source in the $r=0$ axis.

In this paper we will calculate the gravitational energy of
the field configuration considered by Tod, which includes 
altogether disclinations and dislocations. We will obtain the
energy per unit length along the defect axis $r=0$. For a metric
field which describes a disclination only the energy per unit 
length turns out to be {\it exactly} the same of the cosmic 
string. Together with other previous calculations of energy of black
hole configurations, this result supports the validity of 
the present energy expression. 

For the dislocations the result is remarkable. The energy per unit 
length of these defects is zero, when the integration extends
over the whole three dimensional space. This result may be of
significance for cosmology. It may be conjectured that a physical,
cosmological discolation needs a very low energy to be formed, and
hence should play an important role in phase transitions in the
early universe.

The difficulty in obtaining an expression for localized energy
density in the framework of the Hilbert-Einstein action integral 
has led to the widespread belief that the gravitational energy 
cannot be localized. We do not share this idea. 
The energy density of the gravitational field can be naturally
obtained from the Hamiltonian formulation of the teleparallel
equivalent of general relativity (TEGR)\cite{Maluf1}. The TEGR is
an alternative geometrical formulation of Einstein's general 
relativity. The gravitational field in the TEGR is described 
by the tetrad field, and its dynamics is dictated by Einstein's
equations. Therefore this is {\it} not an alternative theory of 
general relativity. The gravitational energy density for 
asymptotically flat geometries has been presented and justified 
in  ref.\cite{Maluf2}.

In section II we present the mathematical preliminaries
of the TEGR, its Hamiltonian formulation and the expression of
the energy for an arbitrary asymptotically flat spacetime. We 
also calculate the energy inside a surface of constant radius
$r_o$ for both the Schwarzschild and the Kerr solutions. In section
III we present the calculation of the energy of conical defects.

\bigskip
\bigskip
\noindent {\bf II. The TEGR in Hamiltonian form}\par
\bigskip

\noindent Notation: spacetime indices $\mu, \nu, ...$ 
and local Lorentz indices
$a, b, ...$ run from 0 to 3. In the 3+1 decomposition latin indices 
from the middle of the alphabet indicate space indices according to
$\mu=0,i,\;\;a=(0),(i)$. The tetrad field $e^a\,_\mu$ and
the spin connection $\omega_{\mu ab}$ yield the usual definitions
of the torsion and curvature tensors:  $R^a\,_{b \mu \nu}=
\partial_\mu \omega_\nu\,^a\,_b +
\omega_\mu\,^a\,_c\omega_\nu\,^c\,_b\,-\,...$,
$T^a\,_{\mu \nu}=\partial_\mu e^a\,_\nu+
\omega_\mu\,^a\,_b\,e^b\,_\nu\,-\,...$. The flat spacetime metric 
is fixed by $\eta_{(0)(0)}=-1$. \\

In the TEGR the tetrad field $e^a\,_\mu$ and the spin connection
$\omega_{\mu ab}$ are completely independent field variables. The 
latter is enforced to satisfy the condition of zero curvature. 
The Lagrangian density in empty spacetime is given by\cite{Maluf1}

$$L(e,\omega,\lambda)\;=\;-ke({1\over 4}T^{abc}T_{abc}\,+\,
{1\over 2}T^{abc}T_{bac}\,-\,T^aT_a)\;+\;
e\lambda^{ab\mu\nu}R_{ab\mu\nu}(\omega)\;.\eqno(1)$$

\noindent where $k={1\over {16\pi G}}$, $G$ is the gravitational 
constant; $e\,=\,det(e^a\,_\mu)$, $\lambda^{ab\mu\nu}$ are 
Lagrange multipliers and $T_a$ is the trace of the torsion tensor
defined by $T_a=T^b\,_{ba}$.   

The equivalence of the TEGR with Einstein's general relativity is         
based on the identity

$$eR(e,\omega)\;=\;eR(e)\,+\,
e({1\over 4}T^{abc}T_{abc}\,
+\,{1\over 2}T^{abc}T_{acb}\,-\,T^aT_a)\,-\,
2\partial_\mu(eT^{\mu})\;,\eqno(2)$$

\noindent which is obtained by just substituting the arbitrary
spin connection $\omega_{\mu ab}\,=\,^o\omega_{\mu ab}(e)\,+\,
K_{\mu ab}$ in the scalar curvature tensor $R(e,\omega)$ in the
left hand side; $^o\omega_{\mu ab}(e)$ is the Levi-Civita 
connection and $K_{\mu ab}\,=\,
{1\over 2}e_a\,^\lambda e_b\,^\nu(T_{\lambda \mu \nu}+
T_{\nu \lambda \mu}-T_{\mu \nu \lambda})$ is the contorsion tensor.
The vanishing of $R^a\,_{b\mu\nu}(\omega)$, which is one of the
field equations derived from (1), implies the equivalence of 
the scalar curvature $R(e)$, constructed out of $e^a\,_\mu$ only, 
and the quadratic combination of the torsion tensor. It also
ensures that the field equation arising from the variation of
$L$ with respect to $e^a\,_\mu$ is strictly equivalent to
Einstein's equations in tetrad form. Let  
${{\delta L}\over{\delta e^{a\mu}}}=0$ denote the field equation 
satisfied by $e^a\,_\mu$. It can be shown by explicit calculations
that

$${{\delta L}\over{\delta e^{a\mu}}}\;=\;{1\over 2}\lbrace R_{a\mu}-
{1\over 2}e_{a\mu}R(e)\rbrace\;.$$

\noindent (we refer the reader to
refs.\cite{Maluf1,Maluf2} for additional details).

It is important to note that for 
asymptoticaly flat spacetimes the total divergence in (2) does
{\it not} contribute to the action integral.  Therefore the latter 
does not require additional surface terms, as it is invariant
under coordinate transformations that preserve the asymptotic 
structure of the field quantities\cite{Faddeev}. 

The Hamiltonian formulation of the TEGR can be successfully 
implemented if we fix the gauge $\omega_{0ab}=0$ from the 
outset, since in this case the constraints (to be 
shown below) constitute a {\it first class} set\cite{Maluf1}.
The condition $\omega_{0ab}=0$ is achieved by breaking the local
Lorentz symmetry of (1). We still make use of the residual time
independent gauge symmetry to fix the usual time gauge condition
$e_{(k)}\,^0\,=\,e_{(0)i}\,=\,0$. Because of $\omega_{0ab}=0$,
$H$ does not depend on $P^{kab}$, the momentum canonically 
conjugated to $\omega_{kab}$. Therefore arbitrary variations of
$L=p\dot q -H$ with respect to $P^{kab}$ yields 
$\dot \omega_{kab}=0$. Thus in view of $\omega_{0ab}=0$, 
$\omega_{kab}$ drops out from our considerations. The above 
gauge fixing can be understood as the fixation of a {\it global}
reference frame.    

Under the above gauge fixing the canonical action integral obtained
from (1) becomes\cite{Maluf1}

$$A_{TL}\;=\;\int d^4x\lbrace \Pi^{(j)k}\dot e_{(j)k}\,-\,H\rbrace\;,
\eqno(3)$$

$$H\;=\;NC\,+\,N^iC_i\,+\,\Sigma_{mn}\Pi^{mn}\,+\,
{1\over {8\pi G}}\partial_k (NeT^k)\,+\,
\partial_k (\Pi^{jk}N_j)\;.\eqno(4)$$

\noindent $N$ and $N^i$ are the lapse and shift functions, 
$\Pi^{mn}=e_{(j)}\,^m\Pi^{(j)n}$  and 
$\Sigma_{mn}=-\Sigma_{nm}$ are Lagrange multipliers. The constraints
are defined by 

$$ C\;=\;\partial_j(2keT^j)\,-\,ke\Sigma^{kij}T_{kij}\,-\,
{1\over {4ke}}(\Pi^{ij}\Pi_{ji}-{1\over 2}\Pi^2)\;,\eqno(5)$$

$$C_k\;=\;-e_{(j)k}\partial_i\Pi^{(j)i}\,-\,
\Pi^{(j)i}T_{(j)ik}\;,\eqno(6)$$

\noindent with $e=det(e_{(j)k})$ and $T^i\,=\,g^{ik}e^{(j)l}T_{(j)lk}$, 
$\;T_{(j)lk}=\partial_l e_{(j)k}-\partial_k e_{(j)l}$.
We remark that (3) and (4) are invariant under {\it global} SO(3) and
general coordinate transformations.  

We assume in this section
the asymptotic behaviour $e_{(j)k}\approx \eta_{jk}+
{1\over 2}h_{jk}({1\over r})$ for $r \rightarrow \infty$. In view
of the relation

$${1\over {8\pi G}}\int d^3x\partial_j(eT^j)\;=\;
{1\over {16\pi G}}\int_S dS_k(\partial_ih_{ik}-\partial_kh_{ii})
\; \equiv \; E_{ADM}\;\eqno(7)$$

\noindent where the surface integral is evaluated for 
$r \rightarrow \infty$, we note that the integral form of 
the Hamiltonian constraint $C=0$ may be rewritten as

$$\int d^3x\biggl\{ ke\Sigma^{kij}T_{kij}+
{1\over {4ke}}(\Pi^{ij}\Pi_{ji}-{1\over 2}\Pi^2)\biggr\}
\;=\;E_{ADM}\;.\eqno(8)$$

\noindent The integration is over the whole three dimensional
space. Given that $\partial_j(eT^j)$ is a scalar  density,
from (7) and (8) we define the gravitational
energy density enclosed by a volume V of the space as\cite{Maluf2}

$$E_g\;=\;{1\over {8\pi G}}\int_V d^3x\partial_j(eT^j)\;.\eqno(9)$$  

\noindent It must be noted that this expression is 
also invariant under global SO(3) transformations. 

We will briefly recall two applications of $E_g$. Let us initially
consider a spherically symmetric geometry and fix the triads 
$e_{(k)i}$ as

$$e_{(k)i}\;=\;\pmatrix{ e^\lambda\,sin\theta cos\phi&r\,cos\theta cos\phi
&-r\, sin\theta sin\phi \cr
e^\lambda\,sin\theta sin\phi & r\,cos\theta sin\phi
& r\, sin\theta cos\phi \cr
e^\lambda\,cos\theta & -r\,sin\theta &0\cr }\eqno(10)$$

\noindent $(k)$ is the line index and $i$ is the column index. The 
function $\lambda(r)$ is determined by

$$e^{-2\lambda}\;=\;1\,-\,{2mG\over r}$$

\noindent The one form  
$e^{(k)}\;=\;e^{(k)}\,_r dr\,+\,e^{(k)}\,_\theta d\theta\,+\,
e^{(k)}\,_\phi d\phi\;$ yields

$$e^{(k)}e_{(k)}\;=\;e^{2\lambda}dr^2\,+\,r^2\,d\theta^2\,+\,
r^2\,sin^2 \theta\, d\phi^2\;.$$

\noindent Therefore the triads given by (10) represent the spatial
section of the Schwarzschild solution. We can easily calculate
$\varepsilon_g = {1\over {8\pi G}}\partial_i(eT^i)$ associated to
(10). We obtain

$$\varepsilon_g\;=\;{1\over G}{\partial \over {\partial r}}
\lbrack r(1-e^{-\lambda})\rbrack\;,\eqno(11)$$

\noindent The energy inside a spherical surface of arbitrary radius
$r_o$ can be calculated from (11). It is given by\cite{Maluf2}

$$E_g\;=\;r_o \biggl\{ 1\,-\,(1-{{2mG}\over r_o})^{1\over 2}
\biggr\}\;.\eqno(12)$$

\noindent This is exactly the expression found by Brown and 
York\cite{Brown} in their analysis of {\it quasi-local} gravitational
energy.  They define a general expression for quasi-local energy
as minus the proper time rate of change of the Hilbert-Einstein
action (with surface terms included), in analogy with the classical
Hamilton-Jacobi equation which expresses the energy of a classical
solution as minus the time rate of change of the action. The 
application of their procedure to the Schwarzschild solution yields
(12). Note that when $ r_o\rightarrow \infty$ we find $E=m$.

The definition (9) for the gravitational energy can also be
successfully applied to the Kerr black hole\cite{Kerr}. In terms
of Boyer and Lindquist coordinates\cite{Boyer} $(t,r,\theta,\phi)$
the spatial section of the Kerr metric is given by

$$ds^2\;=\;{\rho^2 \over \Delta}dr^2\,+\,
\rho^2 d\theta^2\,
+\,{\Sigma^2 \over \rho^2}sin^2\theta d\phi^2 \eqno(13)$$

\noindent with the following definitions:

$$\Delta\; = \;r^2-2mr+a^2\;,$$

$$\rho^2 \; = \; r^2+a^2\,cos^2\theta\;,$$

$$\Sigma^2\;=\;(r^2+a^2)^2\,-\,\Delta\,a^2\,sin^2\theta\;.$$  

\noindent $a$ is the specific angular momentum defined by
$a={J\over m}$. The triads appropriate to the three-metric above
are given by

$$e_{(k)i}\;=\;
\pmatrix{  {\rho \over \sqrt{\Delta}}sin\theta\,cos\phi &
\rho cos\theta\,cos\phi & 
-{\Sigma \over \rho}sin\theta\,sin\phi \cr
{\rho \over \sqrt{\Delta}}sin\theta\,sin\phi &
\rho cos\theta\,sin\phi &
{\Sigma \over \rho}sin\theta\,cos\phi \cr
{\rho \over \sqrt{\Delta}}cos\theta &
-\rho sin\theta& 0 \cr } \eqno(14)$$  

\noindent Indeed, defining again the one-form  
$e^{(k)}\;=\;e^{(k)}\,_rdr\,+\,e^{(k)}\,_\theta d\theta\,+\,
e^{(k)}\,_\phi d\phi\;$
we easily find that $e^{(k)}e_{(k)}=ds^2$ given by (13).

There is another set of triads that yields the Kerr solution, 
namely, the set which is diagonal and whose entries are given by
the squere roots of $g_{ii}$. This set is not appropriate for our 
purposes, and the reason can be understood even in the simple case
of flat spacetime. In the limit when both $a$ and $m$ go to zero
(14) describes flat space: the curvature {\it and} the torsion
tensor vanish in this case. However, for the diagonal set of triads
(again requiring the vanishing of $a$ and $m$),

$$e^{(r)}=dr\;,\;e^{(\theta)}=r\,d\theta\;,\;e^{(\phi)}=
r\,sin\theta\,d\phi\;,$$

\noindent some components of the torsion tensor do not vanish,
$T_{(2)12}=1$, $T_{(3)13}=sin\theta$, and $E_g$ calculated out of
the diagonal set above diverges when integrated over the whole space.
Moreover, it is the flat space form of (14), i.e., when $m=a=0$, that
can be brought to a diagonal form in {\it cartesian} coordinates, 
and not the diagonal form above in spherical coordinates. Thus the 
asymptotic behaviour $e_{(j)k}\approx \eta_{jk}\,
+\,{1\over 2}h({1\over r})$ when $r\rightarrow \infty$ can only
be achieved by means of (14).

In ref.\cite{Maluf3} we have obtained the expression of the 
energy contained within a surface of constant radius $r=r_o$: 

$$E_g\;=\;{1\over 4}\int_0^\pi d\theta\,sin\theta \biggl\{
\rho+{\Sigma \over \rho} - 
{\sqrt{\Delta} \over {\rho \Sigma}}\,\biggl( 2r(r^2+a^2)-
a^2sin^2\theta
(r-m)\, \biggr) \biggr\}_{r=r_o}\;.\eqno(15)$$

In the limit of slow rotation,
namely, when ${a\over r_o}\,<<\,1$ all integrals in the expression
above can be calculated, and $E_g$ finally reads

$$E_g\;=\;r_o\biggl( 1-\sqrt{1-{{2m} \over r_o}+   
{a^2\over r_o^2} }\biggr) 
\;+\;{a^2\over {6r_o}}\,\biggl[ 2+{{2m} \over r_o}+
\biggl( 1+{{2m}\over r_o}\biggr)
\sqrt{1-{{2m} \over r_o}+{a^2\over r_o^2} }
\biggr]\eqno(16)\;,$$

\noindent This is exactly the expression found by 
Martinez\cite{Martinez} who approached the same problem by
means of Brown and York's procedure. However the present 
approach is more general than that of ref.\cite{Martinez}.
The energy given by (15) can be calculated by means of numerical 
integration for any value of $a$. On the other hand Brown and
York's procedure requires the embedding of an arbitrary two
dimensional surface of the Kerr type in the reference space
$E^3$, a construction which is not possible in 
general\cite{Martinez} (the evaluation of the energy in 
ref.\cite{Martinez} is only possible in the limit
${a\over {r_o}}<<1$).
\bigskip
\bigskip

\noindent {\bf III. Conical Defects}\par
\bigskip

The calculations of the previous section
support the correctness of expression (9)
for the energy of the gravitational field. In ref.\cite{Maluf2}
expression (9) was justified in the framework of asymptoticaly
flat gravitational fields. The action integral of the TEGR for
compact spacetimes differs from the one for asymptoticaly flat
geometries by a surface (boundary) term and consequently the two 
Hamiltonian densities also differ by a surface term. However 
the Hamiltonian constraint is the same for both kinds of 
geometries (except, of course, for the possible
presence of {\it additional} 
terms; had we added a cosmological constant to the Lagrangian
density, such a term would also appear in the Hamiltonian 
constraint).

We have seen from (8) that the integral form of the Hamiltonian
constraint equation may be written as $C=H-E_{ADM}=0$
for asymptoticaly flat spacetimes. We will assume here that 
this is a general feature of gravitational theories, namely,
we will assume that for an arbitrary geometry the Hamiltonian
constraint equation may be written in integral form as $C=H-E=0$.
Therefore we will tentatively evaluate the energy
of simple well known geometries which are not asymptoticaly flat
by means of (9). In the following we will evaluate the latter for 
the class of geometries considered by Tod\cite{Tod}. Such a class 
of geometries is particularly suitable for our purposes since the
energy per unit length of a cosmic string is a priori known.

We will consider conical
singularities along the $z$ axis, in an otherwise flat Minkowski
spacetime, described by the metric

$$ds^2\;=\;-(dt+\alpha d\phi)^2+dr^2+{\beta^2}r^2d\phi^2+
(dz+\gamma d\phi)^2\;,\eqno(17)$$

\noindent where $\alpha, \beta$ and $\gamma$ are real constants and
$\phi$ runs from 0 to $2\pi$. The metric is everywhere flat except
at the axis $r=0$. For $\alpha = \gamma =0$ and $\beta \, <\, 1$ the
metric describes a cosmic string type singularity. Thus $\beta$
parametrizes a disclination. As shown in ref.\cite{Tod}, $\alpha$ and
$\gamma$ parametrize dislocations, in the terminology used in
crystallography. The metric as given above has also been considered
by Gal'tsov and Letelier\cite{Galtsov}. The spatial section of (17) 
reads

$$g_{ij}\;=\;
\pmatrix { 1&0&0\cr
0&{\delta}^2&\gamma\cr
0&\gamma&1\cr }\eqno(18)$$

\noindent where $\delta^2 \equiv \beta^2r^2+\gamma^2-\alpha^2$. The 
corresponding triads are given by

$$e_{(k)j}\;=\;
\pmatrix { cos\phi & -\delta sin\phi & 
-{\gamma \over \delta}sin\phi\cr
sin\phi & \delta \cos\phi & 
{\gamma \over \delta}cos\phi\cr
0& 0& \sqrt{1-{\gamma^2 \over \delta^2}}\cr } \eqno(19)$$

Recall that $(k)$ and $j$  are the line and column index, 
respectively. Note that if we make $\alpha = \gamma =0\;,\,\beta =1$
(19) can be brought to a diagonal form in cartesian coordinates by
means of a coordinate transformation.
Initially we evaluate the components of the torsion tensor:

$$T_{(1)12}\;=\;{\partial \over {\partial r}}\biggl(-\delta\,sin\phi
\biggr)\,-\,{\partial \over {\partial \phi}}\biggl(cos\phi \biggr)\;
=\;(1-\delta ')sin\phi$$

$$T_{(1)13}\;=\;{\partial \over {\partial r}}\biggl(
-{\gamma \over \delta}sin\phi \biggr)\,
-\,{\partial \over{\partial z}}\biggl(cos\phi \biggr)\;
=\;{\gamma \over \delta^2}\,\delta ' \,sin\phi$$

$$T_{(1)23}\;=\;{\partial \over {\partial \phi}}\biggl(
-{\gamma \over \delta}sin\phi \biggr)\,-\, 
{\partial \over {\partial z}} \biggl( -\delta\,sin\phi \biggr)\;
=\;-{\gamma \over \delta} cos\phi$$

$$T_{(2)12}\;=\;-(1-\delta')cos\phi$$

$$T_{(2)13}\;=\;-{\gamma \over \delta^2}\,\delta' cos\phi$$

$$T_{(2)23}\;=\;-{\gamma \over \delta}sin\phi$$

$$T_{(3)12}\;=\;0$$

$$T_{(3)13}\;=\;{\delta'\over \delta}\,{ { \gamma^2\over\delta^2} \over
\sqrt{1-{\gamma^2\over\delta^2}} }$$

$$T_{(3)23}\;=\;0\;,$$

\noindent where the prime denotes differentiation with respect to $r$.
We wish to evaluate (9) and for this purpose we need the expression
of $T^i$. After a long but simple calculation we obtain

$$T^1\;=\;{1\over \delta}(1-\delta')\,
-\,{\gamma^2 \over {\delta^2-\gamma^2}}\,{\delta' \over \delta}\;,$$

$$T^2\;=\;T^3\;=\;0\;.$$

\noindent Together with $e=\sqrt{\delta^2-\gamma^2}$, the energy 
density can now be easily obtained and integrated. The energy 
$E_g$  contained within a cylindrical region with length $L$ 
and radius $r_o$ is given by

$$E_g\;=\;{L\over 4}\,\sqrt{1-{\gamma^2 \over \delta^2}}\,
\biggl\{ 1- \biggl( {\delta^2 \over {\delta^2 -\gamma^2}}\biggr) \,
{\beta^2 r_o \over \delta} \biggr\} \eqno(20)\;.$$

We will consider next the three individual situations in which 
the metrics are parametrized by only one of the parameters.\par

\vskip 2.5cm
\noindent {\bf I.} $\;\;\;\alpha=\gamma=0$\par

\noindent The metric parametrized by $\beta$ only describes 
a disclination. Expression (20) reduces to 

$$E_g \;=\;{L\over 4}(1-\beta)\eqno(21)\;.$$

\noindent This is precisely the energy per unit lenght for a cosmic
string\cite{Vilenkin}. We note that $E_g$ above does not depend on
the radius of integration $r_o$. Therefore we may conclude that the
whole energy is concentrated along the defect axis $r=0$. \par

\bigskip
\noindent {\bf II.} $\;\;\;\alpha=0,\;\beta=1$\par
\noindent In this case we have a simple dislocation parametrized by
$\gamma$. We are mostly interested in the value of $E_g$ 
for very large values of $r_o$.  From (20) we obtain

$$E_g\;\approx\;-{\gamma^2 \over 8}\,{L\over {r_o}^2}\;.\eqno(22)$$

\noindent Therefore in the limit when both $L$ and $r_o$ go to
infinity, namely, when the integration is performed over the whole 
three dimensional space, $E_g$ vanishes. Thus the total energy of 
in the disclocation is zero. However, in the limit $r_o \rightarrow 0$ 
we find $E_g \rightarrow -{L\over 4}$.\par

\bigskip
\noindent {\bf III.} $\;\;\;\beta=1, \gamma=0$\par
\noindent For this metric expression (20) reduces to

$$E_g\;=\; {L\over 4}\biggl( 1- {1\over \sqrt{1-{\alpha^2 \over {r_o}^2}}}
\biggr)\;.$$

\noindent For large values of $r_o$ we obtain

$$E_g\; \approx \;-{\alpha^2 \over 8}\,{L\over {r_o}^2} \eqno(23)\;.$$

\noindent Therefore the total energy corresponding to this
disclination is also zero.  In the limit $r_o \rightarrow \alpha$
we observe that $E_g \rightarrow -\infty$.

We observe that whereas for a disclination the whole energy
per unit length is concentrated along the defect axis, for 
both types of dislocations the energy is distributed over 
the whole  three dimensional space. 

\bigskip
\bigskip
\noindent {\bf IV. Comments}\par
\bigskip

By means of expression (9) for the energy of the 
gravitational field we have
obtained the correct value of the energy per unit length of a
cosmic string. This fact is a good indication that the Hamiltonian
constraint may be generically written as $C=H-E=0$, which is the
assumption we made in this paper. It also supports the conjecture
that expression (9) might have a universal character, since it
yields the expected values of energy for totally distinct 
spacetimes, namely, Kerr-type and conical spacetimes.

From the results of section {\bf III}
we conclude that dislocations are more likely to
appear as a result of cosmological phase transitions in the
early universe  than disclinations (cosmic strings). The energy
of an actual, physical dislocation might be nonvanishing, but 
anyhow we would expect it to be very small. The situation here
is very much similar to what happens in a real crystal. It is well
known that disclinations require too much energy to be formed 
in crystals, whereas dislocations are more favourable 
defects since they require much less energy (see, for instance,
sections 6.3.2 and 6.5 of ref.\cite{Rivier} for a discussion as 
to why the energy cost for a disclination in crystal is 
prohibitively high). In order to understand the vanishing of the 
total gravitational energy for dislocations, 
let us consider the metric for which $\alpha=0,
\beta=1$ (case II). As discussed by Tod\cite{Tod}, this metric
can be transformed into a flat metric if we define a new 
coordinate $Z=z+Z_o{\phi \over {2\pi}}=z+\gamma \phi$. The
associated Burgers vector is determined by $Z_o=2\pi \gamma$.
We know, however,  that the energy of an actual dislocation
in a crystal depends not only on the Burgers vector but
also on an extra quantity,  the rigidity modulus $\mu$.
Such quantity is not given in (17). This fact might explain why
the total energy of the dislocations above is zero.
Disclinations and dislocations are concepts used in the 
deformation models of crystals and metals. We conclude from
our analysis that dislocations of type considered here might
be as well useful concepts in cosmological models.\par

\bigskip
\noindent {\it Acknowledgements}\par
\noindent This work was supported in part by CNPQ.


\newpage


\begin{thebibliography}{99}


\bibitem{Vilenkin}
A. Vilenkin, Phys. Rep. {\bf 121}, 265 (1985)

\bibitem{Zeldovich}
Ya. B. Zel'dovich, Mon. Not. R. Astr. Soc. {\bf 192}, 663 (1980)

\bibitem{AVilenkin}
A. Vilenkin, Phys. Rev. Lett. {\bf 46} 1169 (1981); 1496 (1981)


\bibitem{Nabarro}
F. R. N. Nabarro, {\it Theory of Crystal Dislocations} (Dover,
New York, 1967) 


\bibitem{Rivier}
N. Rivier, in {\it Geometry in Condensed Matter 
Physics}, Ed. J. F. Sadoc (World Scientific, Singapure, 1990)

\bibitem{Tod}
K. P. Tod, Class. Quantum Grav. {\bf 11}, 1331 (1994)

\bibitem{Vickers}
J. A. G. Vickers, Class. Quantum Grav. {\bf 4}, 1 (1987)

\bibitem{Maluf1}
J. W. Maluf, J. Math. Phys. {\bf 35}, 335 (1994)

\bibitem{Maluf2}
J. W. Maluf, J. Math. Phys. {\bf 36}, 4242 (1995)

\bibitem{Faddeev}
L. D. Faddeev, Sov. Phys. Usp. {\bf 25}, 130 (1982)


\bibitem{Brown}
J. David Brown and J. W. York, Jr, Phys. Rev. {\bf D47}, 1407 (1993)


\bibitem{Kerr}
R. P. Kerr, Phys. Rev. Lett. {\bf 11}, 237 (1963)

\bibitem{Boyer}
R. H. Boyer and R. W. Lindquist, J. Math. Phys. {\bf 8}, 265 (1967)

\bibitem{Maluf3}
J. W. Maluf, E. F. Martins and A.Kneip,  {\it Gravitational energy of
rotating black holes}, Univ. de Bras\'ilia preprint (1996) (submitted
to the J. Math. Phys.).




\bibitem{Martinez}
E. A. Martinez, Phys. Rev. {\bf D50}, 4920 (1994)

\bibitem{Galtsov}
D. V. Gal'tsov and P. S. Letelier, Phys. Rev. {\bf D47}, 4273 (1993)



\end{thebibliography}
\end{document}